\documentclass[useAMS]{mn2e}
\usepackage{amsmath}
\usepackage{textcomp}
\usepackage{amssymb}
\usepackage[pdftex]{graphicx}
\usepackage{amssymb}

\usepackage[margin=.8in, paperwidth=8.5in, paperheight=11in]{geometry}
\usepackage{multirow}
\usepackage{multicol}
\usepackage{epsfig}
\usepackage{graphicx}
\usepackage{esint}
\begin{document}

\title{Line Driven Acceleration using Multi-Frequency Radiation Hydrodynamics}
\author[S. Dyda et al.]
{\parbox{\textwidth}{Sergei~Dyda$^{1}$\thanks{sdyda@ast.cam.ac.uk}, Christopher S. Reynolds$^{1}$, Yan-Fei Jiang$^{2}$}\\
$^{1}$Institute of Astronomy, Madingley Road, Cambridge CB3 0HA, UK \\
$^{2}$Center for Computational Astrophysics, Flatiron Institute, New York, NY 10010, USA
}

\date{\today}
\pagerange{\pageref{firstpage}--\pageref{lastpage}}
\pubyear{2018}

\label{firstpage}

\maketitle

\begin{abstract}
We use multi-frequency radiation hydrodynamics (rad-HD) to simulate radiative acceleration of a spherically symmetric stellar wind. We demonstrate the rad-HD capabilities of \textsc{Athena++} for a series of test problems with multi-group radiation transfer. We then model the radiative transfer of a single spectral line through a spherically symmetric, isothermal, ``CAK''-type line driven wind. We find that correctly accounting for the Doppler shift of the absorbed radiation, the force is well described by the analytic Sobolev line transfer in the supersonic parts of the solution where the flow is stationary and the effects of Abbott waves is negligible. Unlike in the analytic, steady-state solution re-radiation is important and leads to non-trivial radiation energy density and fluxes in the outer parts of the wind. We discuss a variety of applications to these multi-group methods that are currently computationally tractable.      
\end{abstract}


\section{Introduction}
Astronomical observations make use of the full electromagnetic spectrum, though hydrodynamic modeling efforts of multi-frequency radiation, owing primarily to computational expense, have thus far been lacking. This, despite the fact that many multi-frequency radiation methods are available for a variety of hydrodynamics codes. For example, Kuiper et al. (2010) implemented a multi-frequency solver  using flux limited diffusion (FLD) in the magnetohydrodynamics (MHD) code \textsc{PLUTO}. Rosen et al. (2017) developed a multi-frequency  hybrid  radiation  hydrodynamics  module  that  adaptively combines long characteristics and a moment method for the GRMHD code \textsc{HARM}$^2$. Gonz\'alez et al. (2015) developed a multi-group radiation solver using FLD for \textsc{RAMSES}. Pawlik \& Schaye (2011) developed a multi-frequency radiation method for the smooth-particle hydrodynamics (SPH) code \textsc{GADGET}. Higgenbottom et al. (2018 and references therein) coupled a MCMC radiation code \textsc{PYTHON} to the MHD code \textsc{ZEUS}. 

\textsc{Athena++} (Stone et al. in prep) is a \textsc{C++} rewrite of the MHD code \textsc{Athena} (Gardiner \& Stone 2005, 2008) with flexible coordinate systems and adaptive-mesh refinement, improved scalability and new physics such as general relativity (White, Stone \& Gammie 2016) and radiation transport (Jiang, Stone \& Davis 2012, 2014 , hereafter JSD14). Subsequent versions of this code have improved on JSD14, where radiation terms were accurate to $\mathcal{O} \sim (v/c)$, by converting the specific intensity between the lab and co-moving frame, where the hydro and radiation terms are computed respectively (Jiang, Stone \& Davis 2019, hereafte JSD19). Here we extend this algorithm to allow multiple frequencies and frequency dependent scattering and absorption cross sections coupled via Doppler shifts. We first test the code using multi-group methods, where the different frequencies represent radiation bands widely separated in frequency space and effects like Doppler or turbulent broadening may be ignored. We then develop a method accounting for the Doppler shifting of frequencies when this condition is relaxed. After testing, we apply it to studying radiation propagating in a spherically symmetric, isothermal, ``CAK''-type line driven wind in frequencies near a single optically thick spectral line. 

Line driving is a promising mechanism for explaining the acceleration of flows from a variety of astrophysical objects - massive stars, cataclysmic variables (CVs) and active galactic nuclei (AGN). Lucy and Soloman (1970, hereafter LS70) showed that in the context of massive stars the presence of hundreds of optically thick lines at the base of stellar atmospheres could provide an enhancement to the radiative pressure above that due to electron scattering alone. This could allow the radiative force to overcome gravity even in the case of a sub-Eddington source. Castor, Abbot \& Klein (1975, hereafter CAK) then showed that crucially, if the flow is optically thin to the continuum and accelerating fast enough, then optically thick lines will be Doppler shifted and continue being accelerated by continuum photons, the so called Sobolev approximation. The condition on the flow acceleration is described by the optical depth parameter
\begin{equation}
t = \frac{\rho v_{\rm{th}} \sigma_e}{dv/dr},
\label{eq:sobolev}
\end{equation}     
where $\rho$ is the gas density, $v_{\rm{th}}$ the gas thermal velocity, $\sigma_e$ the electron scattering cross section and $dv/dr$ the acceleration per unit distance in the flow. Physically, the optical depth parameter compares the width of a line to its Doppler shift after traveling a photon mean free path through the flow. The optical depth parameter $t$ determines the force multiplier $M(t)$, which characterizes the effective number of optically thick lines available for radiative acceleration. The value at which the force multiplier saturates, $M_{\rm{max}}$ determines the threshold luminosity required to overcome the inward force of gravity which is approximately

\begin{equation}
L_{\rm{min}} \sim \frac{L_{\rm{Edd}}}{M_{\rm{max}}},
\end{equation}
i.e. by what factor below the Eddington luminosity $L_{\rm{max}}$ can winds still be launched. Phenomenologically, line driving has been successful on a variety of fronts - predicting mass flux and outflow velocities from massive stars, variablity of emission line profiles from O stars and absorption and emission profiles from CVs and AGN. Given these broad successes it is critical to carefully verify the fundamental assumptions behind the model, such as the validity of the Sobolev approximation and the dynamics of the line transfer, to formulate testable predictions for observers. 

One line of attack has been to carefully study the propagation of radiation through the flow. Semi-analytic treatments were used to study radiation transfer in spherically symmetric flows (Kunasz \& Hummer 1974a,b, Mihalas, Kunasz \& Hummer 1975). Later analyses have shown that line driven winds are unstable. Abbott (1980) showed that the CAK solution is unstable to density perturbations in the flow. Later 1D simulations by Owocki (1984) showed that these perturbations can grow and produce density features, so called clumps, on \emph{sub}-Sobolev length scales, though re-radiation of the line may help stabilize the flow (Lucy 1984). Further simulations showed that these instabilities persisted in 2D (Dessart \& Owocki 2005; Sundqvist et al. 2018). Others have relaxed the idealized assumption of a point source and used a star of finite angular extent, which leads to a reduced mass loss rate (Friend \& Abbott 1986; Pauldrach, Puls \& Kudritzki 1986).  

Another approach has been to revisit the microphysics describing the interaction between the radiation field and the gas. Studies using photoionization codes have improved on the initial estimates of LS70 for the number of optically thick lines (Gayley 1995; Puls et al. 2000). Others have corrected the line force due to changes in the ionization state of a spherical flow (Abbott 1982). Recently, Dannen et al. (2018) have investigated wind models which relax equation (\ref{eq:sobolev}) because photoionization studies show that different parts of the wind are dominated by different ionic species.

In this work, we devote our computational resources to propagate photons of different frequencies through a gas to study radiatively driven acceleration. Section \ref{sec:numerical} describes our code and basic numerical setup. In Section \ref{sec:tests} we describe tests of our multi-group numerical methods for cases with and without Doppler shifting. In Section \ref{sec:results} we present results for a model of a ``CAK''-type line driven wind where multiple frequencies around a single optically thick line is propagated through the flow, which we compare to analytic results predicted by Sobolev theory. In Section \ref{sec:discussion} we discuss possible applications of these methods for studying line driven winds.  We conclude in Section \ref{sec:conclusion} where we comment on future applications of multi-group rad-HD simulations that are already computationally tracktable or will be in the near future.

\section{Numerical Methods}
\label{sec:numerical}

We performed all numerical simulations with the developmental version of the rad-MHD code \textsc{Athena++} (JSD14 for the main numerical methods and JSD19 for the latest updates). The numerical tests in Section \ref{sec:tests} use a 2D box in pressure equilibrium. The box contains one or two high density spherical clouds. Radiation flux enters the box along a fixed direction, which is assumed to be emitted from a distant radiation source. We study either the radiation transport (for problems where we keep the hydrodynamics fixed) or cloud acceleration via radiation pressure. In Section \ref{sec:results} we study the acceleration of a spherically symmetric, isothermal, line driven wind in spherical polar coordinates in 1D.  We describe the basic equations of rad-hydro in Section \ref{sec:setup} and describe our algorithm for accounting for Doppler shifting in \ref{sec:doppler_alg}.

\subsection{Basic Equations}
\label{sec:setup}
In dimensionless form the basic equations for single fluid hydrodynamics coupled to a radiation field are
\begin{subequations}
\begin{equation}
\frac{\partial \rho}{\partial t} + \nabla \cdot \left( \rho \mathbf{v} \right) = 0,
\end{equation}
\begin{equation}
\frac{\partial (\rho \mathbf{v})}{\partial t} + \nabla \cdot \left(\rho \mathbf{vv} + \mathsf{P} \right) = -\mathbb{P} \mathbf{S_r}(\mathbf{\mathbf{P}}),
\label{eq:momentum}
\end{equation}
\begin{equation}
\frac{\partial E}{\partial t} + \nabla \cdot \left( (E + P)\mathbf{v} \right) = - \mathbb{PC} S_r(E),
\label{eq:energy}
\end{equation}
\label{eq:hydro}%
\end{subequations}
where $\rho$, $\mathbf{v}$ are the fluid density and velocity respectively and $\mathsf{P}$ is a diagonal tensor with components P the gas pressure. The total gas energy is $E = \frac{1}{2} \rho |\mathbf{v}|^2 + \mathcal{E}$ where $\mathcal{E} =  P/(\gamma -1)$ is the internal energy and $\gamma = 1.01$. The isothermal sound speed is $a^2 = P/\rho$ and the adiabatic sound speed $c_s^2 = \gamma a^2$. The temperature is $T = (\gamma -1)\mathcal{E}\mu m_{\rm{p}}/\rho k_{\rm{b}}$ where $\mu = 1.0$ is the mean molecular weight and other symbols have their standard meaning. The absorption and scattering cross sections are $\sigma_a$ and $\sigma_s$. We define the dimensionless radiation pressure $\mathbb{P} = P_0/a_rT_0^4$ and speed of light $\mathbb{C} = c /a_0$ where the $0$ subscript denotes fiducial values in the problem and $a_r$ is the radiation energy density constant. Unless otherwise indicated we use $\mathbb{P} = 10^{-2}$ and $\mathbb{C} = 2.1 \times 10^{3}$. The radiation source terms $\mathbf{S_r}$ and $S_r(E)$ are calculated for each frequency by the differences between the angular quadratures of the specific intensity $I(n)$ in the lab frame before and after adding the source terms (see JSD19). Radiation moments of the angular quadrature over all the solid angles $\Omega$ are then defined as
\begin{subequations}
\begin{align}
J_{\nu} = \int I_{\nu} \, d\Omega,
\end{align}
\begin{align}
\mathbf{H}_{\nu} = \int \mathbf{n} \, I_{\nu} \, d\Omega,
\end{align}
\begin{align}
\mathsf{K}_{\nu} = \int \mathbf{n}\mathbf{n} \, I_{\nu} \, d\Omega.
\end{align}
\end{subequations}
The frequency dependent moments are related to the frequency dependent radiation energy density $E_{r,\nu}$, flux $\mathbf{F}_{r,\nu}$ and pressure $\mathsf{P}_{r,\nu}$ via $E_{r,\nu} = 4 \pi J_{\nu}$, $\mathbf{F}_{r,\nu} = 4 \pi c \mathbf{H}_{\nu}$ and $\mathsf{P}_{r,\nu} = 4 \pi \mathsf{K}_{\nu}$. Naturally we define the total radiation density $E_r = \int d\nu E_{r,\nu}$ and other quantities likewise.

\subsection{Doppler Shift}
\label{sec:doppler_alg}

Scattering and absorption opacities are implemented in \textsc{Athena++} in the rest frame of the gas. Transforming from the lab to the co-moving frame, frequencies Doppler shift according to 
\begin{equation}
\nu_0 = \Gamma \left( 1 - \frac{\mathbf{v} \cdot \mathbf{n}}{c}\right) \nu,
\label{eq:doppler} 
\end{equation}
where $\Gamma = 1/\sqrt{1 - (v/c)^2}$ is the usual relativistic factor. Using the Lorentz invariant intensity (see for example Mihalas \& Mihalas 1984) one can write
\begin{equation}
I(\nu_0) \ d \nu_0 = \left( \frac{\nu_0}{\nu}\right)^4 I(\nu) \ d \nu.
\label{eq:intensity}
\end{equation}  
At every time-step, the code transforms intensity from the lab frame to the co-moving frame, applies radiation source terms (involving scattering and absorption opacities) and then converts back to the lab frame intensity. If the width of frequency bands are large compared to gas velocity $\Delta \nu \gg v/c \ \nu_0$, we can neglect the Doppler shift resulting from (\ref{eq:doppler}) and assume the argument of the intensity is unchanged in (\ref{eq:intensity}). This is the approximation used in the multi-group implementation of the code. If this assumption does not hold, as is the case when modeling a line profile, we must account for the Doppler shift when transforming between frames.    

When initializing the problem we define a frequency grid with $N_{\nu}$ bins of width $\Delta f$ over the range $f_0 \leq \nu \leq f_0 + N_{\nu} \Delta f$. We specify the frequency dependent intensity $I(\nu)$ and the frequency dependent scattering and absorption opacities in the rest-frame of the gas. At every time-step, we apply (\ref{eq:doppler}) to each frequency bin of the lab frame intensity array and use cubic interpolation to compute the intensity at each discrete frequency. Details of this algorithm can be found in Appendix \ref{sec:appendix}. We then compute source terms as before, update the intensity array in the co-moving frame, before inverting (\ref{eq:doppler}) and interpolating between frequencies to convert back to the lab frame. For computational purposes we assume that Doppler shifts are periodic in frequency space - physically this assumption is reasonable provided the Doppler shifts expected in our simulation are small relative to the size of the grid. To maintain the consistency of (\ref{eq:doppler}) when shifting between frames under this periodic assumption, we approximate $\nu \approx f_0$, i.e the frequency shift is identical for all frequencies, which is a good approximation provided $N_{\nu} \Delta f \ll f_0$.

\section{Multi-Frequency Tests}
\label{sec:tests}

We use various models to test the multi-frequency capabilities of \textsc{Athena++}. We divide these into multi-group models with no Doppler shift and $N_{\nu} = 2$ frequencies, meant to simulate bands widely separated in frequency space and models with Doppler shift and $N_{\nu} \geq 30$, meant to simulate line profiles or other phenomena where frequencies are narrowly separated relative to typical velocities in the problem.

\subsection{Initial Conditions}
\label{sec:ic}

To study the interaction between matter and radiation, our setup consists of a 2D box of gas in hydrostatic equilibrium. In some setup we include ``clouds", consisting of circular over-dense regions of gas with density profile 
\begin{equation}
\rho = \rho_0 + \frac{\rho_1 - \rho_0}{1 + \exp(10(r -1))},
\end{equation}
where $\rho_1 $ is the maximum cloud density and $r = [(x - \Delta x)/x_0]^2 + [(y - \Delta y)/y_0]^2$. Here $x_0 = y_0$ is the radius of the cloud and $(\Delta x,\Delta y)$ the coordinates of the cloud center. Because the higher density cloud is in pressure equilibrium with the ambient gas, its temperature is less than $T_0$.

On the top and right sides of the box we impose outflow conditions on the gas variables and vacuum conditions on the radiation.  Along the bottom and left side of the box we keep density and pressure kept fixed at $\rho_0$ and $P_0$ respectively, while ensuring velocity is conserved when we perform this update.

We use this setup because the direction of rays $\hat{n}_i$ does not align with any coordinate directions. By choosing $n_{\rm{ang}} = 4$ radiation rays and allowing radiation to enter from the left and bottom parts of the box, radiation is directly incident on the cloud in the coordinate system rotated to lie along $\hat{n}_1$. This is effectively a 1D problem, but we are correctly capturing the shadowing effect of the cloud. All our tests use this setup in 2D, but we perform an effectively 1D analysis along $\hat{n}_1$. 

\subsection{No Doppler Shift Models}
We perform simulations where we model two frequency bands denoted $\nu_1$ and $\nu_2$. We test the propagation of radiation in a domain with static gas configuration (Section \ref{sec:camel}) and the acceleration of an optically thin cloud (Section \ref{sec:delta_acceleration}). 

\subsubsection{Camel Test}
\label{sec:camel}

\begin{figure}
                \centering
                \includegraphics[width=0.48\textwidth]{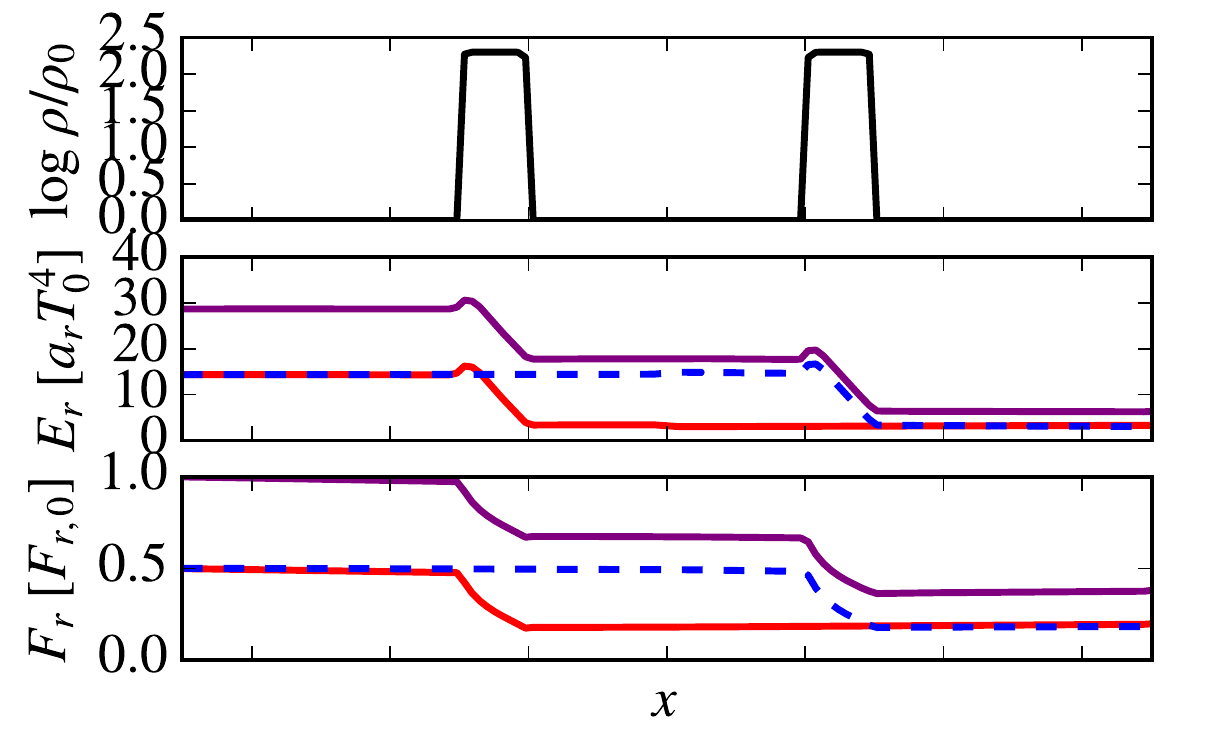}
        \caption{1D slice for camel test along the line $y = x$ passing through the center of the clouds. We plot the density (top panel), radiation energy density (middle panel) and radiation flux (bottom panel), indicating contributions from $\nu_1$ (red), $\nu_2$ (blue) and the total (purple).}
\label{fig:camel_1d}
\end{figure} 

We propagate the radiation field through two clouds of radius $r_0 = 0.05$ centered at $(\Delta x, \Delta y) = (\pm 0.25, \pm0.25)$ and density $\rho = 200 \rho_0$. We keep the clouds fixed by reinitializing the initial conditions at every full time-step, via the user workin loop. The goal of this test is to ensure that the radiation transfer leads to the same steady state for each photon frequency and captures the relevant ``shadowing" behind their respective optically thick cloud. The gas opacity is given by
\begin{equation}
\sigma = \begin{cases} 
      \sigma_0 \ \delta(\nu - \nu_1) & x \leq 0 \\
      \sigma_0 \ \delta(\nu - \nu_2) & x > 0 \\
   \end{cases}
\end{equation}
i.e frequency $\nu_1$ interacts with cloud \# 1 and frequency $\nu_2$ interacts with cloud \#2.

In Fig \ref{fig:camel_1d} we plot the density $\rho$ (top panel), the radiation energy $E_r$ (middle) and flux $F_r$ (bottom panel) for frequencies $\nu_1$ (red), $\nu_2$ (blue) as well as the total (purple) along the 1D slice $y = x$ passing through the center of the cloud. This corresponds to a steady state for the radiation. The energy flux of each band decreases after passing through their respective optically thick cloud, with the exception of the cloud interface where re-radiation leads to a slight increase. The parallel radiation flux decreases almost linearly through the cloud and remains constant in each half of the simulation domain where the gas is optically thin.  As expected, both frequencies behave identically with their respective cloud with non-zero opacity.

\subsubsection{Delta-function Line Acceleration}
\label{sec:delta_acceleration}

\begin{table*}
\begin{center}
    \begin{tabular}{| l | c | c | c| c |c| c | c | c|c|c|c|c|c|c|}
    \hline \hline
	&		&\multicolumn{2}{|c|}{Intensity} & \multicolumn{2}{|c|}{Opacity} & Acceleration & \multicolumn{6}{|c|}{Flux} \\
Model	& $N_{\gamma}$ 	& $I_1$	&$I_2$ 	& $\sigma_1$	&$\sigma_2$ 	& $a \ [\times 10^{-4}]$&$F_{\rm{in}}$ & $F_{\rm{out}}$& $F_{\nu_1,\rm{in}}$ &$F_{\nu_1,\rm{out}}$ &$F_{\nu_2,\rm{in}}$ &$F_{\nu_2,\rm{out}}$    \\ \hline \hline
$N_1$	& 1 		& 90.50	&- 	& 0.1		&- 		& 16.2 &18.5 &15.1  &18.5  &15.1 &-    & -    \\ \hline
$N_{11}$& 2 		& 45.25	&45.25 	& 0.1		&0.1 		& 16.2 &18.5 &15.1  &9.2   &7.6  &9.2  & 7.6  \\
$N_{01}$& 2 		& 45.25	&45.25 	& 0.0		&0.1 		& 8.1  &18.5 &16.8  &9.2   &9.2  &9.2  & 7.6   \\
$N_{10}$& 2 		& 45.25	&45.25  & 0.1		&0.0 		& 8.1  &18.5 &16.8  &9.2   &7.6  &9.2  & 9.2   \\ \hline \hline  
    \end{tabular}
\end{center}
\caption{Summary of $N_{\nu} = 2$ simulations with delta-function absorption lines. We indicate the scattering cross sections $\sigma_{i}$ for frequency $\nu_i$, the cloud acceleration $a$ and the radiation flux entering $F_{\rm{in}}$ and exiting $F_{\rm{out}}$ the simulation domain for each frequency as well as the total. As expected, sharing the total radiation flux amongst two frequency bands results in the same cloud acceleration (model $N_{11}$) whereas making the cloud optically thin to one of the bands results in half the acceleration (models $N_{01}$ and $N_{10}$).}
\label{tab:summary}
\end{table*}  

We test cloud acceleration in a pure scattering regime using two frequencies with an optically thin cloud. The goal of this test is to show that treating the radiation as two groups with the same scattering properties leads to the same dynamics as using a single frequency and grey opacity. Likewise, using two groups but making the cloud optically thin to one of the frequencies, halves the cloud acceleration. We take the opacity
\begin{equation}
\sigma = \sigma_1 \delta(\nu - \nu_1) + \sigma_2 \delta(\nu - \nu_2). 
\end{equation}
i.e where spectral lines are modeled by delta-functions. We take the central cloud density $\rho_1 = 10 \rho_0$. This corresponds to the simplest case ``S10" in Proga et al. (2014). The cloud experiences a uniform acceleration, due to the radiation flux attenuated by the optical depth of the cloud. The flux exiting the cloud can be approximated by $F_{\rm{out}} = e^{-2\tau} F_{\rm{in}}$, where $\tau = 2 x_0 \sigma$ is the optical depth of the cloud and $F_{\rm{in}}$ the incident radiation flux. From the momentum equation, the cloud acceleration $a = \mathbb{P} F \sigma$. Substituting our expression for the flux and integrating over the thickness of the cloud we find the total acceleration
\begin{equation}
a = \frac{\mathbb{P} F_{\rm{in}}}{2} \left( 1 - e^{-2 \tau}\right). 
\label{eq:a_rad}
\end{equation}
A summary of our models is listed in Table {\ref{tab:summary}}, where we list the incoming and outgoing radiation flux in each frequency band and the resulting cloud acceleration. In the single frequency case $N_1$, with intensity $I_0 = 90.50$ and opacity $\sigma_0 = 0.1$, the acceleration $a \approx 16.7 \times 10^{-4}$, comparable to the acceleration observed in simulations of $a \approx 16.2 \times 10^{-4}$. 

We consider the following extensions of this experiment using two frequencies. In all cases we keep the total flux constant by halving the intensity of each frequency band $I_1 = I_2 = I_0/2$. In model $N_{11}$ the scattering cross section in each frequency is kept constant, $\sigma_1 = \sigma_2 = \sigma_0$. The cloud behaves as in the fiducial case $N_1$, accelerating at the same rate, which is expected since the physics is identical except we are now modeling two physically identical frequency bands. Further both frequencies behave identically as far as their transmission through the cloud. The models $N_{10}$ ($N_{01}$) keep the same total flux, but the scattering cross section of frequency $\nu_1$ ($\nu_2$) is set to zero. The flux incident on the cloud that can provide a radiation force is thus halved, leading to an acceleration $a \approx 8.1 \times 10^{-4}$,  half that of the fiducial case. The $N_{10}$ and $N_{01}$ cases otherwise behave symetrically with respect to exchanging $\nu_1$ and $\nu_2$.

\subsection{Doppler Shift Models}
We perform simulations where we model $N_{\nu} = 30$ frequencies, equally spaced in the band $f_0 \leq f \leq f_0 + N_{\nu} \Delta f$ where $f_0 = 1000$ and $\Delta f = 0.02$. In Section \ref{sec:doppler} we study the Doppler shifting of the absorption profile of a static gas cloud moving relative to the radiation field source. In Section \ref{sec:line_acceleration} we compare the acceleration of a cloud due to scattering from a spectrally resolved line and from greybody scattering.     

\subsubsection{Doppler Shifted Line Absorption}
\label{sec:doppler}

\begin{figure}
                \centering
                \includegraphics[width=0.48\textwidth]{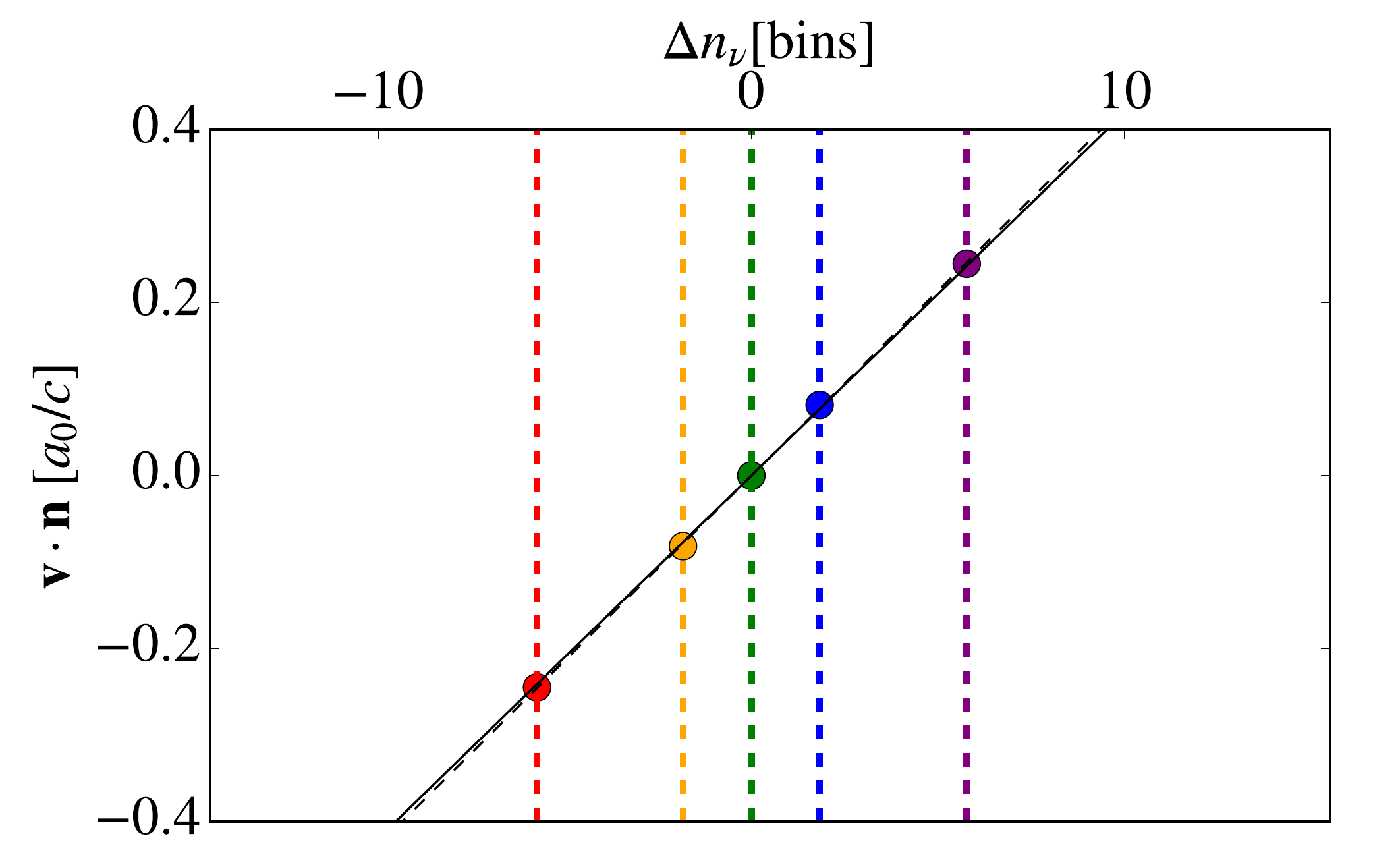}
                \includegraphics[width=0.48\textwidth]{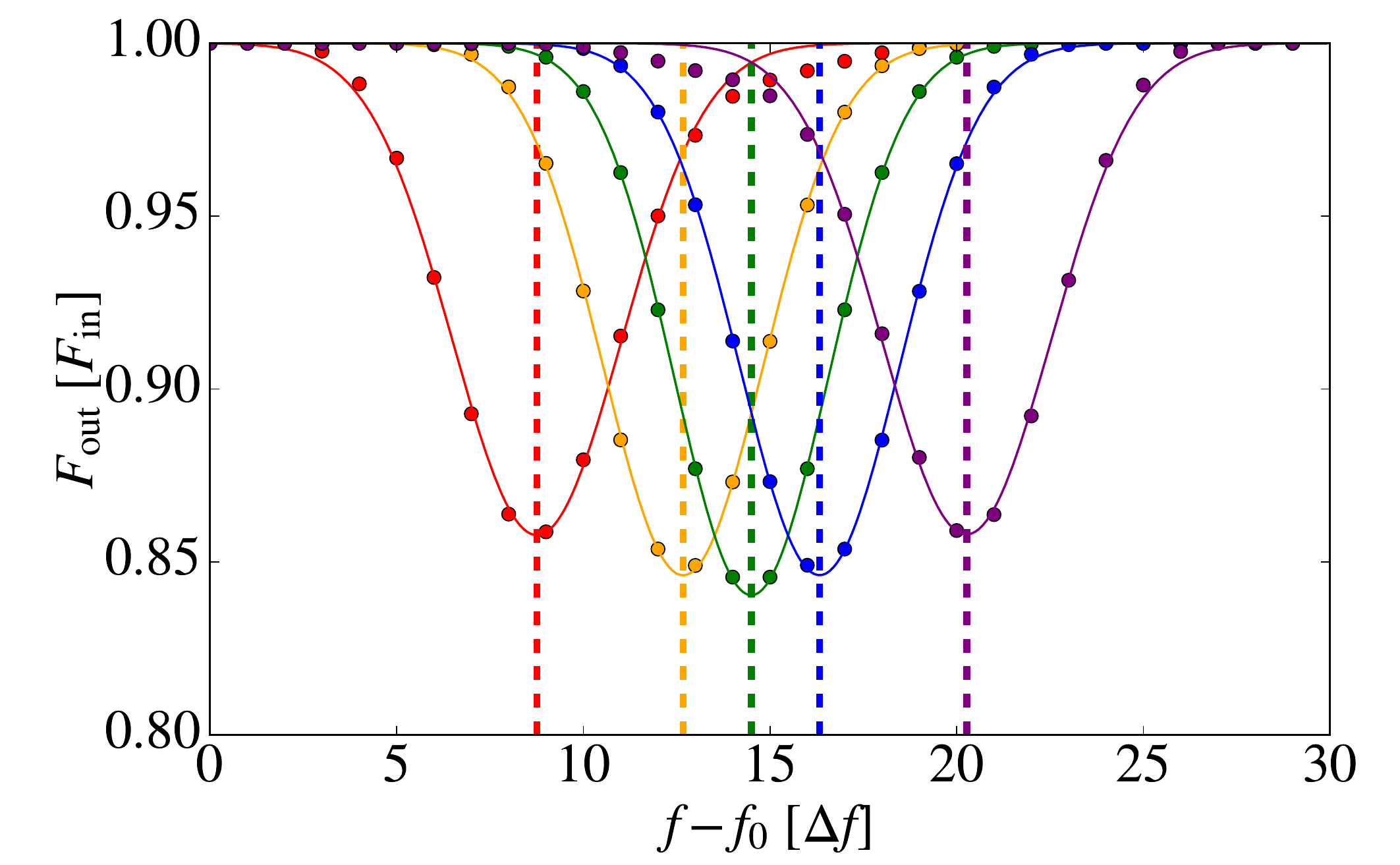}
        \caption{\textit{Top -}Doppler shift in the absorption profile minimum in units of frequency bins $n_{\nu}$ and frequency $\Delta \nu$ for clouds with $v = $ -0.3 (red), -0.1 (orange), 0 (green), 0.1 (blue) and 0.3 (purple). The expected shift $\Delta \nu = f_0 v/c$ is shown with the dashed line. \textit{Bottom -} Doppler shifted absorption profiles for the same cases (colored points) and the corresponding Gaussian fit (colored lines). The shifts are symmetric with respect to the sign of $\mathbf{v}$ but larger velocities lead to a broadening of the trough.}
\label{fig:dopler_shift}
\end{figure} 

\begin{figure}
                \centering
                \includegraphics[width=0.48\textwidth]{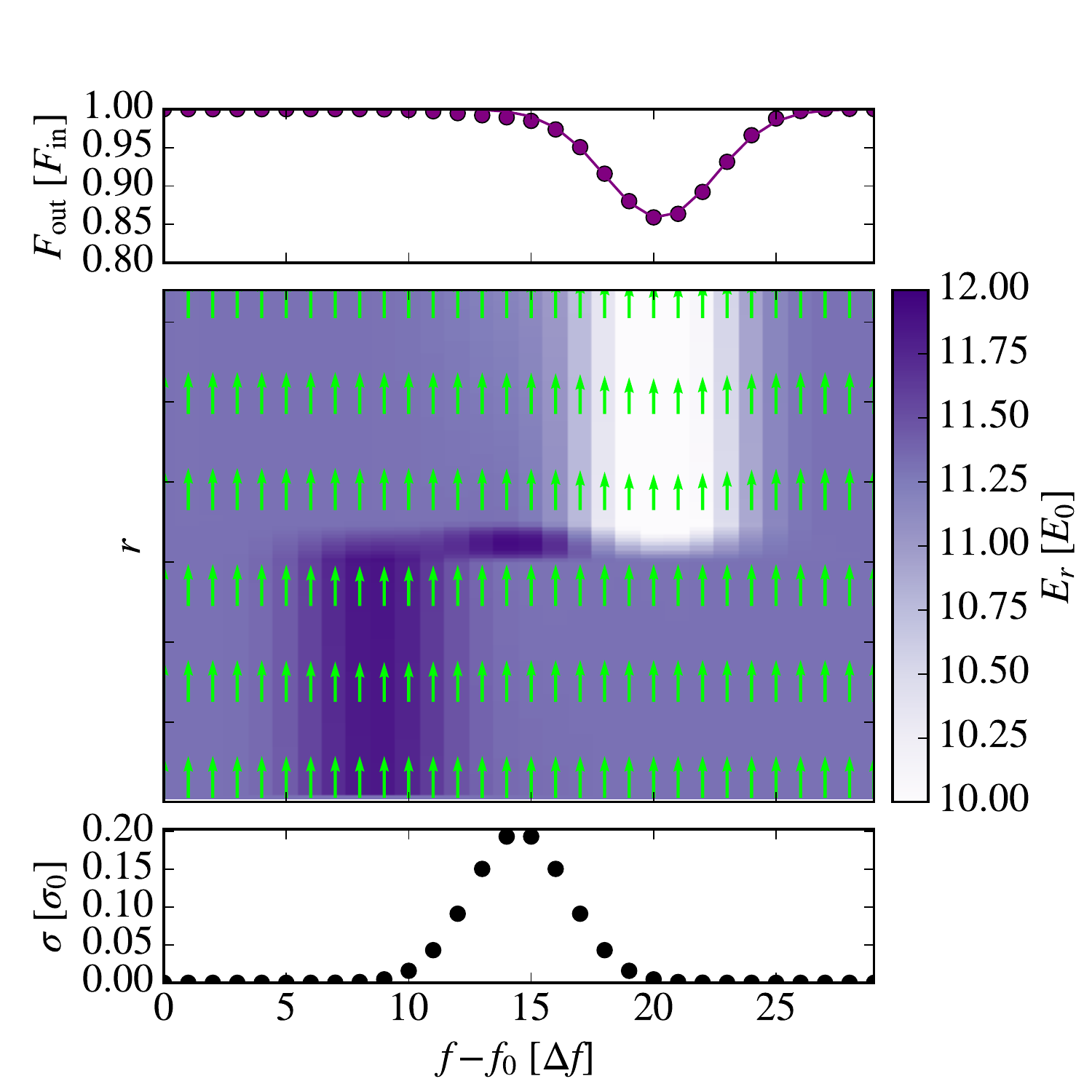}
        \caption{Doppler shifted line for $v = 0.3$ for $N_{\nu} = 30$. The cloud primarily absorbs radiation in higher frequency bands due to the cloud moving away from the radiation source. Re-radiated light is emitted in the cloud rest frame, and Doppler shifted to lower frequencies. \textit{Top panels} - Flux exiting the outer boundary as a function of frequency, normalized by the input flux. \textit{Middle panels} - Radiation energy density (color) and flux (vectors) as a function of radius for each frequency. \textit{Bottom panels} - Frequency dependent rest frame scattering opacity.}
\label{fig:spectrum}
\end{figure} 

We test this setup by irradiating a cloud moving at constant velocity relative to the radiation field. We consider a model where we assume a single optically thick, thermally broadened line with opacity
\begin{equation}
\sigma_s(f) = \sigma_0 \frac{\Delta f}{\sqrt{2 \pi w^2_{\sigma}}} \exp \left\{ -\frac{1}{2}\left(\frac{f - f_L}{w_{\sigma}}\right)^2\right\},
\label{eq:line_opacity}
\end{equation}
where $f_L = f_0 + (N_{\nu} - 1)\Delta f/2$ is line center and the width of the line $w_{\sigma} = 2 \Delta f$ . We set the normalization by requiring $\int df \sigma_s(f) = \sigma_0$. We only test the radiation transport of the code, by re-initializing the hydro variables to their initial values after every time-step.  

We compute the transmitted flux through the cloud as a function of cloud frequency. In the lower panel of Fig \ref{fig:dopler_shift} we plot the Doppler shifted absorption line profile (colored points) for clouds with velocity $v = $ -0.3 (red), -0.1 (orange), 0 (green), 0.1 (blue) and 0.3 (purple) and fit each of these profiles to a Gaussian (colored lines) and extract the line center. In the upper panel we plot the shift in line center as a function of velocity projected onto the direction of incident radiation, $\mathbf{v} \cdot \mathbf{n}$ for each of the above profiles in the corresponding color. The fit for the Doppler shifted profiles is shown with the solid line, which we compare to the theoretical curve (dashed line), generated from (\ref{eq:doppler}). We find good agreement between the two in this case and note the absorption profiles retain their Gaussian shape. We find some dispersion of the profile, due to resolution in frequency space. For the v = 0 case and $N_{\nu} = 30$ we find a fractional change in the profile width $\Delta \sigma_w/\sigma_w = 3\%$. When $v = 0.3$ this fractional change increases to $\Delta \sigma_w/\sigma_w = 13\%$. By comparison, using a linear interpolator, rather than a cubic interpolation as we have done leads to deviations $\Delta \sigma_w/\sigma_w = 50\%$ for $v = 0.3$.  If we increase the resolution to $N_{\nu} = 100$ frequency bins, the fractional change decreases to $\Delta \sigma_w/\sigma_w = 8\%$. 

The Gaussian line profile shape is maintained even when the periodicity of the frequency grid come into play. We used $N_{\nu} = 30$ frequencies, in which case for the range of velocities explored, periodicity of the frequency grid only affects bins outside the core of the line profile. We emphasize that in less idealized problems however it may be hard to correctly track all frequency bins and therefore a suitably large grid should be chosen if computationally possible.

In Fig. \ref{fig:spectrum} we plot the flux exiting the outer boundary as a function of frequency, normalized by the input flux (top panel), the radiation energy density (color) and flux (vectors) as a function of radius for each frequency (middle panel) and frequency dependent rest frame scattering opacity (bottom panel) for the case $v = 0.3$ (purple case in Fig. \ref{fig:dopler_shift}). It illustrates subtle effects due to re-radiation. The cloud is moving away from the radiating source, so the transmission of bluer frequencies in suppressed. Because the cloud is a rigid body, any absorption of radiation re-emitted within the cloud occurs at line center. Hence we see an enhancement in the energy density inside the cloud around $f_L$. Finally, in the rest frame of the cloud, gas closer to the radiation source is receding away from the cloud. Therefore radiation from this gas is red shifted i.e the term $\mathbf{v} \cdot \mathbf{n}$ has flipped sign causing an enhancement to the energy density in the red part of the spectrum. This re-radiation effect may be important for capturing instabilities in line driven winds (see for example Lucy 1984).      

We conclude that radiative effects are complex, even in the most simple and contrived experiments. For this reason, we will err on the side of caution and include sufficient freqeuncy grid resolution so that periodicity of the frequency grid does not come into play for the gas velocities we explore.           

\subsubsection{Spectrally Resolved Line Acceleration}
\label{sec:line_acceleration}
As a new application of this method we model the acceleration of a cloud via radiation pressure due to scattering by a line modeled via (\ref{eq:line_opacity}). As a benchmark, we compare the dynamics to a single frequency model with greybody opacity. For $\sigma_0 = 0.01$, $\mathbb{P} = 10^{-2}$, $\rho_1 = 10$ and $I_0 = 45.25$ we expect from (\ref{eq:a_rad}) an acceleration $a = 9.11 \times 10^{-4}$ and find $a = 9.33 \times 10^{-4}$ for both the $N_{\nu} = 1$ greybody opacity case and the $N_{\nu} = 20$ case where we model the line profile. This is an optically thin case where we expect such agreement to hold. As opacity is increased, the line center will be optically thick, whereas the edges will remain optically thin. Further we have explicitly turned off Doppler shifting effects, though for some choices of parameters the Doppler shift cannot be neglected. Exploring this case more fully will be key left for a future study.

\section{Radiation Driven Stellar Wind}
\label{sec:results}
We study a spherically symmetric, radiation driven wind where we explicitly treat the radiation transfer through the wind. Our starting point is the CAK (1975) analytic solution for a line driven wind. CAK assumes the wind is optically thin to a central source of continuum radiation with Eddington fraction $\Gamma_*$ but experiences a radial force 
\begin{equation}
\mathbf{F}_{\rm{rad}} = \Gamma_* \left(1 + M(t) \right) \frac{GM}{r^2} \hat{r},
\label{eq:rad_force}
\end{equation}   
where the first term in the brackets is due to electron scattering and second due to radiation pressure due to lines, described by the force multiplier  
\begin{equation}
M(t) = k t ^{-\alpha},
\end{equation}
and the optical depth parameter is given by (\ref{eq:sobolev}). In this work we resolve the radiation transfer of a single line through a wind that is primarily driven by a radiation force given by (\ref{eq:rad_force}). We use this line to verify the validity of the radiation transfer approximations in CAK and establish the effects of fully accounting for its effects on the wind structure. We test our setup by first explicitly treating the radiation force due to electron scattering using full radiation transfer (Section \ref{sec:ee_scatter}). We then model the radiation pressure due to a single spectral line (Section \ref{sec:line_scatter}) through a wind launched by a radiation force given by (\ref{eq:rad_force}).

We use typical parameters for a stellar line driven wind (see Dyda \& Proga 2018) but convert to dimensionless parameters where lengths scale with gravitational radii $r_g = GM/c^2$ and velocities to the speed of sound $a_0$. In these units the central potential $GM = \mathbb{C}^2 = 3.13 \times 10^{10} r_g^3 s^{-2}$, $\rho_* = 3.16 \times 10^{5} \ g r_g^{-3}$, $v_{\rm{th}} = 2.48\ a_0$, $r_* = 3.13 \times 10^{7} r_g$, $\sigma_e = 1.808 \times 10^{-11} \ g r_g^{2}$. We take the gas to radiation pressure ratio $\mathbb{P} = 10^{-3}$ and the dimensionless speed of light $\mathbb{C} = 1.7707 \times 10^{5}$.  For a fixed Eddington fraction $\Gamma_* = 0.1$, we find a steady-state solution with constant mass flux $\dot{M} = 4.60 \times 10^{18} g/s$, in agreement with the analytic solution of CAK.

\subsection{Electron Scattering}   
\label{sec:ee_scatter}

\begin{figure}
                \centering
                \includegraphics[width=0.45\textwidth]{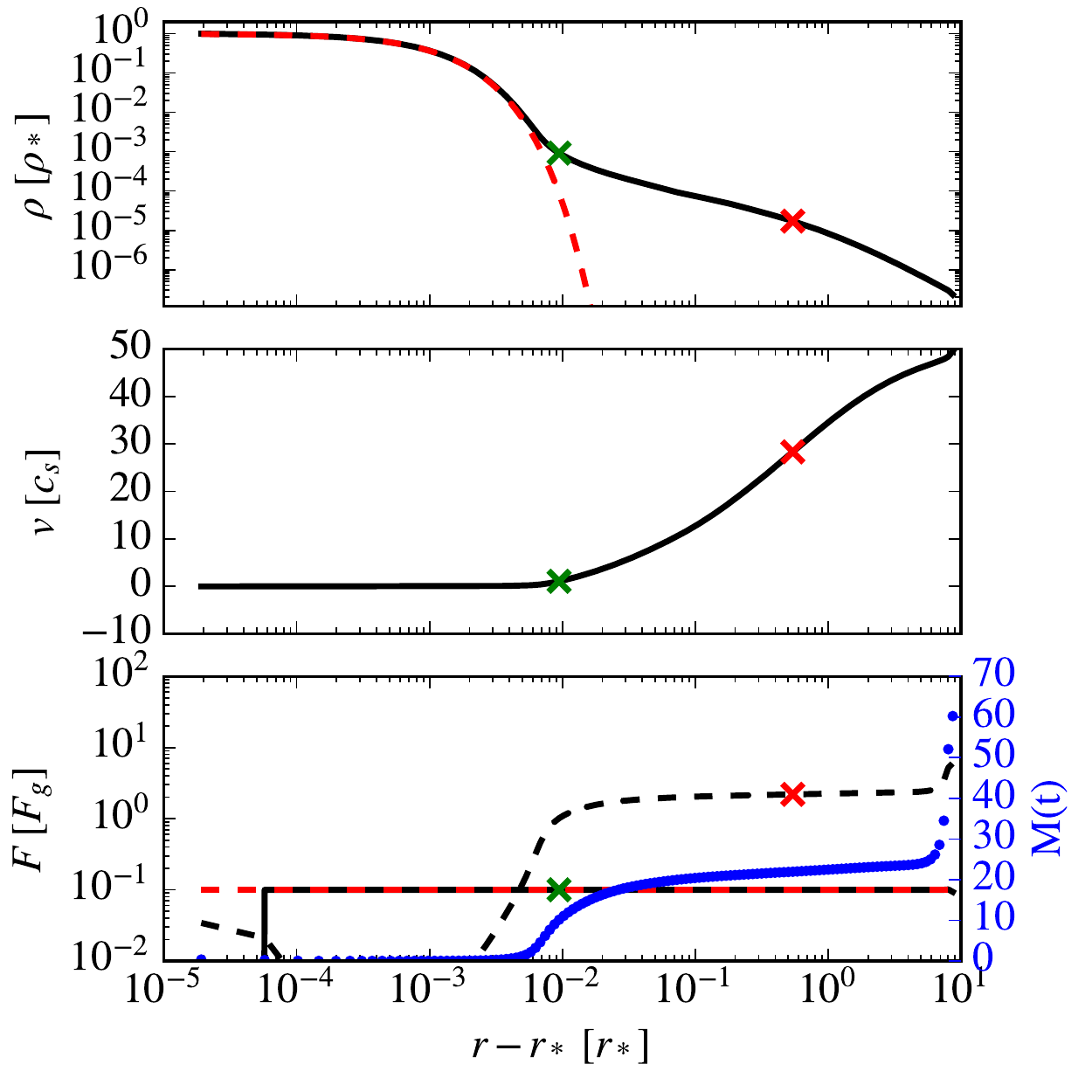}
        \caption{Dynamical variables as a function of radius for radiation driven wind where we model $N_{\nu} = 60$ frequency bins and the corresponding coupling to the gas via electron scattering. We indicate the sonic point with a green X and the critical point with a red X. \textit{Top -} Density follows an exponential atmosphere, indicated by the dashed red line at small radii before assuming a $\sim r^{-2}$ profile of a stationary wind. \textit{Middle -} The velocity flow is monotonically increasing \textit{Bottom -} Forces acting on the flow, scaled to the local gravitational force, including the radiation force due to line driving (dashed black line) and electron scattering (solid black line). The latter is in good agreement with the same force in the optically thin limit (red dashed line) given by the first term in (\ref{eq:rad_force}). We also show the value of the force multiplier (blue dots).}
\label{fig:ee_scatter}
\end{figure} 

We explicitly compute the force due to electron scattering by using a grey opacity $\sigma_e = 1.808 \times 10^{-11}$ $g r_g^{2}$ and $N_{\nu} = 60$ frequencies and intensity $I = 1.0422 \times 10^{7}$. The radiation force is given by (\ref{eq:rad_force}) where the term in brackets is now simply $M(t)$.

In Fig. \ref{fig:ee_scatter} we plot the density $\rho$ (top panel), velocity $v$ (middle panel) and forces (bottom panel) for this solution in the steady state. The density and velocity distributions show a smooth transition from a nearly static, exponential atmosphere, shown as a dashed red line. The velocity crosses the critical point (red cross), where $dv/dr = v/r$ far beyond the sonic point (green cross), since thermal energy is negligible in this flow. The force plot shows that the flow is primarily driven by the radiation force due to spectral lines (black dashed line), which becomes dominant to gravity at or beyond the sonic point. Importantly, from our perspective, the radiation force due to electron scattering (solid black line) well reproduces the force due to electron scattering in the optically thin limit $F_e = \Gamma_* GM/r^2$ (dashed red line), with both lines overlapping except at the first calculated cell. This is as expected since for electron scattering the wind optical depth $\int \rho \sigma_e dr \sim 10^{-5} \ll 1$. At the outermost part of the flow the force multiplier increases sharply. This is because the velocity distribution becomes steeper, hence $dv/dr$ increases, the optical depth parameter t increases leading to an increase in $M(t)$ (blue points). 

\subsection{Pressure due to a Single Line}
\label{sec:line_scatter}

\begin{figure}
                \centering
                \includegraphics[width=0.45\textwidth]{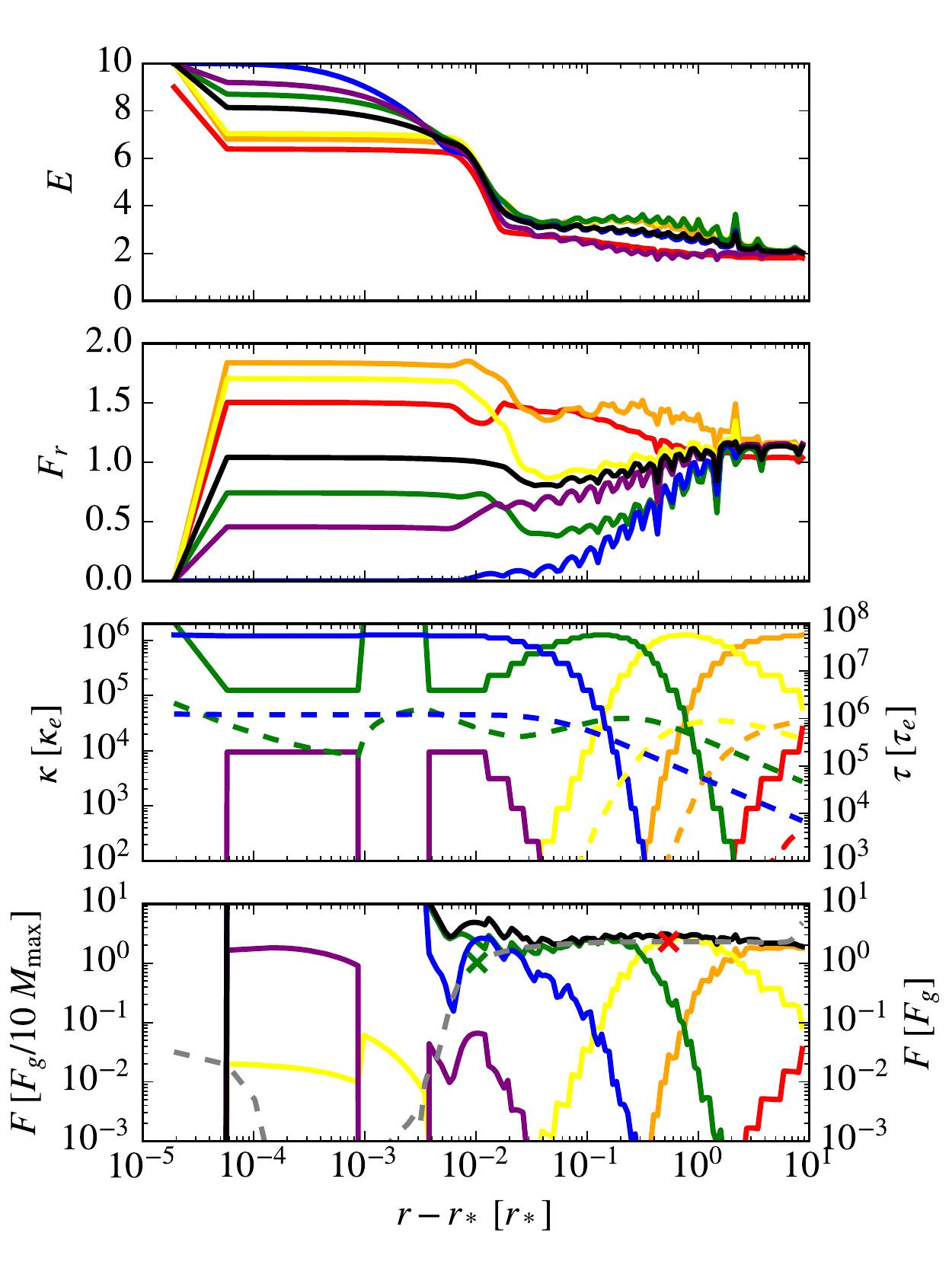}
        \caption{Radiation variables as a function of radius for radiation driven wind where we model $N_{\nu} = 60$ frequency bins and the corresponding coupling to the gas via a single spectral line. We indicate the sonic point with a green X and the critical point with a red X in the bottom panel. Each of the colored lines represents the total contribution from n = 10 frequency bins. \textit{Top -} Radiation energy density $E_r$ \textit{Second -} Radiation flux $F_r$ \textit{Third -} Opacity $\kappa$ (solid lines) and optical depth $\tau$ (dashed lines) for each of the frequency bands. Different parts of the flow are optically thick to different frequencies, depending on the flow velocity. \textit{Bottom -} CAK line driving force acting on the flow, scaled to the local gravitational force, in the optically thin limit (dashed grey line). The colored lines are the force due to different frequency bands, scaled to $10 M_{\rm{max}}$ times the local gravity and the solid black line the total force due to all frequency bands. This shows that the radiation force of the single modeled line provides some fraction of the total line force in the optically thin limit.}
\label{fig:ld_scatter}
\end{figure} 

We model the radiative transfer of a single line through this accelerating flow by propagating $N_{\nu} = 60$ frequencies through the wind. The frequency dependent opacity
\begin{equation}
\sigma_{L}(n) =  \frac{\eta \sigma_e}{\sqrt{2 \pi \sigma_w^2}} \exp\left\{-\frac{1}{2}\left(\frac{6n - 4N_{\nu} - 3}{4 \sigma_w}\right)^2 \right\},
\label{eq:line_profile}
\end{equation}  
where $\eta = \kappa_L/\sigma_e$ is a re-scaling of the line opacity in terms of the electron scattering cross section, $\sigma_w = 2 \Delta f$ controls the line profile width and we have chosen to center the line profile to peak between frequency bins $40 \leq n \leq 41$. We assume that frequencies are Doppler shifted via (\ref{eq:doppler}) with $f_0 = 10^{15} Hz$ and use a frequency grid spacing $\Delta f = 10^{10} Hz$. For the velocities in our unperturbed simulation, these frequencies yield Doppler shifts of $\sim 30$ bins in frequency space.   

For purposes of our analysis, we divide the $N_{\nu} = 60$ frequency bins into six bands with 10 frequencies each: $1 \leq n_{\nu} \leq 5$ and $56 \leq n_{\nu} \leq 60$ (red), $6 \leq n_{\nu} \leq 15$ (yellow), $16 \leq n_{\nu} \leq 25$ (orange), $26 \leq n_{\nu} \leq 35$ (green), $36 \leq n_{\nu} \leq 45$ (blue) and $46 \leq n_{\nu} \leq 55$ (purple). A static atmosphere is therefore optically thick to the blue band and optically thin to the others. For the expected Doppler shifts of $\lesssim 30$ bins, the wind is optically thin to the red and purple bands throughout the wind but the other bands will see different parts of the wind as optically thick. We will also sometimes plot frequency averaged quantities (black).

The dynamical variables are as in the solution shown in Fig. \ref{fig:ee_scatter}. In Fig. \ref{fig:ld_scatter} we plot the radiation energy density (top panel), flux (second panel), opacity and optical depth (third panel) and forces on the wind (fourth panel). 

We see a qualitatively different behaviour for frequency bands near line-center and far from line-center. The blue band has an enhanced energy density at the base of the wind by a factor of $\sim 2$ relative to the bands away from the line (red, orange, etc..). The blue band flux is nearly zero at the base of the wind, but increases steadily as the opacity decreases due to Doppler shifting. We see similar, but less pronounced effects for the green and purple bands that are near line center but still experience non-trivial opacity effects at the base of the wind where velocity is low. Contrast this to the bands that see an optically thin wind base (red, orange, yellow). These have constant flux over the small radial range of the wind base and only begin to drop off when the opacity becomes non-negligible.

We note that re-radiation effects are important in this solution. We ran a case where the Doppler shift effect was turned off. In this case, frequencies near line-center were quickly absorbed and the flux went to zero. The blue band flux vanished near $r - r_* \sim 10^{-2} \ r_*$. The optically thin bands however decreased as $F \sim (r/r_*)^{-2}$ as expected due to geometric dilution i.e. they behaved precisely as the bands in the pure electron scattering case. Here however we see that the flux at the outer radius is roughly constant for all energy bands, as non-zero opacity allows the radiation to be scattered. A constant flux solution is expected, since we have evolved the wind to a steady state. In a dynamical solution, we may expect a different result and leave the exploration of such solutions to later work.    

The third panel shows the lab frame opacity $\kappa$ (solid lines) and opacity $\tau$ (dashed lines). At small radii, where the velocity is low, the opacity is high for the blue band (around which the line rest-frame is centered). As the flow accelerates, the opacity becomes dominated by the green, yellow and orange bands respectively. The optical depth increases until the opacity peaks, after which it begins to decrease because of our choice of normalization i.e. After $\kappa$ reaches a maximum, the rate of increase in the line opacity decreases relative to the rate of increase of electron scattering that has constant $\kappa_e$.

The fourth panel shows the magnitude of the driving force in units of the local gravitational force (dashed grey line). We also plot the force due to each frequency band (colored lines) and the total force due to all frequency bins (solid black line), scaled to $\sim 10 M_{\rm{max}}$. With this empirical scaling, we show that the force due to a single modeled line provides a fraction of the line force in the optically thin limit from the CAK approximation. The two methods disagree at the very base of the wind, where velocity and velocity gradients are small and the optical depth parameter is highly variable. This is unsurprising as in the subsonic part of the flow even the mass outflow rate is variable to to small scale fluctuations (Abbott 1980).  The force due to any radiation band is proportional to the opacity, so we see different parts of the wind being accelerated by the dominant contribution to the line opacity, as expected.

\section{Discussion}
\label{sec:discussion}

We note that we have modeled a slightly broader line than predicted by our choice of gas thermal velocity. This was done to reduce the number of frequency points required to to resolve the entire range of Doppler shifted frequencies. For thermal velocities considered in CAK the FWHM of a line due to thermal broadening is approximately
\begin{equation}
\Delta f_{\rm{FWHM}} \approx \frac{v_{\rm{th}}}{c} \sqrt{8 \ln 2} f_0.
\end{equation}   
Directly from (\ref{eq:line_profile}), $\Delta f_{\rm{FWHM}} \approx 8 \Delta f$. Our line profile is therefore $\sim 8$ times broader than predicted from thermal broadening alone. We have done this for computational simplicity, since otherwise we would have required a larger frequency grid to accomodate Doppler shifts in the entire wind.

We have modeled a single line to understand the propagation of a single radiation frequency and primarily driven the outflow using the CAK mechanism. An alternative model is to assume a distribution of lines and assume the total line force $F_{\rm{rad}} \sim \hat{n} \cdot \nabla(\hat{n} \cdot \mathbf{v})$. This was the formalism used by Kee et al. (2016) to study line-driven ablation of circumstellar discs in Oe/Be stars (see Kee (2015) for in in-depth description of their numerical methods). Such a treatment is possible using our code, which would allow us to capture effects from multipe resonance points, which are ubiquitous for non-spherical disc winds. 

Other studies have used iterative schemes to find solutions to the radiative transfer problem. Earlier models used a simplified form for the radiation tranfer to model the hydrodynamics, such as the usual Sobolev approximation and CAK prescription. Detailed multi-frequency radiation transfer models, including micro-physics such as non-LTE and multi-line scattering and co-moving frame radiation line transfer that forgoes the Sobolev approximation, were then used to compute line-profiles consistent with this prescribed hydrodynamics (see for example Lobel \& Blomme 2008; Hennicker et al. 2018). The current state of the art is coupling the radiation transfer and hydrodynamics codes to solve for a fully self-consistent stationary solution, rather than relying on a simplified treatment of the radiation transfer during the hydrodynamics calculation. Such studies have shown that stellar mass loss rates are highly dependent of microphysics, such as Sander, Vink \& Hamann (2019, hereafter SVH19) who showed $\dot{M}$ depends on metalicity. The inadequacy of the CAK formalism was recognized early on, in particular the need to treat the line driving parameters $k$ and $\alpha$ as variable (see Kudritzki 2002). This can be shown explicitly a posteriori using these self-consistent radiation hydrodynamics solutions (see for example SVH19, to make this inadequacy apparent. Recognizing these inadequacies of the CAK formalism in predicting mass loss rates in massive stars, we have nonetheless used it to generate a simple wind solution to carry out our multi-frequency radiation transfer. Given this simplified treatment of the radiation force, we found that the radiative transfer of a single Gaussian line profile is consistent with what we expect from CAK. 

Since the theoretical work of Abbott (1980) and simulations of Owocki \& Rybicki (1984) it has been known that line driven winds are unstable, due to the line deshadowing instability (LDI). The LDI is responsible for generating sub-Sobolev length structure in the base of the wind and may contribute to the growth of larger clumps in the wind. In this work we devote computational resources to resolving the transition of the flow through the critical point. Our grid resolution is formally larger than the Sobolev length, so we cannot capture such instabilities. Dedicated simulations capturing sub-Sobolev lengths would be required, in a domain smaller than the radial range over which the wind accelerates to near terminal velocity. Understanding the formation of such density features may be important in properly inferring the mass loss rates of OB stars. Effects involving the LDI using multi-frequency rad-HD, such as scattered radiation for instance, is left to future investigations.

\section{Future Work}
\label{sec:conclusion}

With growing computational capabilities we will be able to investigate frequency dependent effects in hydrodynamics. Such effects have already begun to reveal themselves as important for well known problems in astrophysics. For instance, Takeo et al (2019) investigated black hole accetion and showed that the critical mass required for accretion flows to transition to a super-Eddington regime is different when irradiated by a non-powerlaw spectrum because higher mass black holes have harder spectra an over-ionize the surrounding gas. Likewise, in the context of outflows, Huang, Davis \& Zhang (2019) have shown that AGN clouds can be efficiently accelerated by radiation but tend to be dissipated if the UV to IR flux ratio is close to or greater than unity (2019).    

Multi-band methods with small numbers of bands add little in terms of computational cost in comparison to greybody rad-HD, but allow modeling of completely new phenomena. For example, we may be in a position to model coronal heating in X-ray binaries, with a low frequency band for blackbody disc photons and high energy coronal photons. We may construct more accurate models of thermally driven winds (Higginbottom et al. 2018) as photoionization codes have demonsrated that the heating/cooling rates are highly dependent on the incident SED, which in turn affects the wind launching (Dyda et al. 2017). 

As demonstarted in the case of re-radiation from a Doppler shifted cloud (Section \ref{sec:doppler}), radiative effects can be subtle. This suggests that new observational signatures may be found by computing radiative effects \textit{ab initio}, rather than in post-processing. For example, in the case of AGN clouds, Waters et al. (2017) showed in post-processing, that comparison of the peaks of an OVII doublet can be used as a diagnostic of the cloud acceleration. With the advent of high resolution X-ray spectroscopy such as ARCUS, XRISM or Athena which may resolve such phenomena, the need for high fidelity simulations which capture radiative physics becomes important.

Similarly we may begin modeling radiative effects such as line locking (Arav 1995), where emission features appear in NV and CIV BALs. Such simulations require modeling only a few spectral lines (in this case NV, CIV and the Ly-$\alpha$ emission) and is computationally feasible in 2D. The line locking mechanism explicitly requires re-emission to occur, so simulations involving full radiation transfer should be informative. 

\section*{Acknowledgements}
S.D. would like to thank Jim Stone for stimulating discussions on line driving and Zhuahuan Zhu, Daniel Proga and the Physics \& Astronomy department at UNLV for graciously hosting the Athena++ workshop where this work was started. S.D. acknowledges support from ERC Advanced Grant 340442. C.S.R. thanks the UK Science and Technology Facilities Council (STFC) for support under the New Applicant grant ST/R000867/1, and the European Research Council (ERC) for support under the European Union's Horizon 2020 research and innovation programme (grant 834203). The Center for Computational Astrophysics at the Flatiron Institute is supported by the Simons Foundation.

\appendix
\section{Doppler Shifting Algorithm}
\label{sec:appendix}

\begin{figure}
                \centering
                \includegraphics[width=0.48\textwidth]{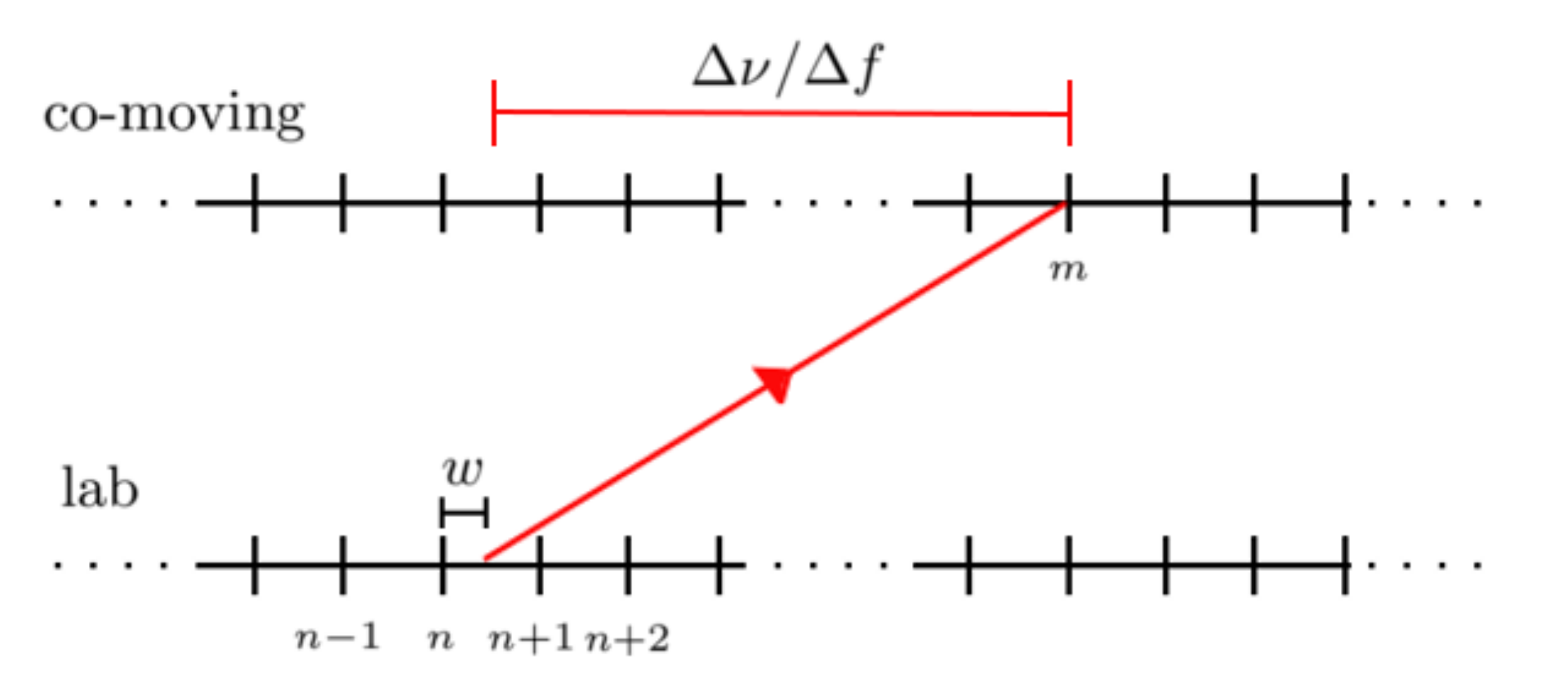}
        \caption{Diagram illustrating the change in frequency indexing of the radiation intensity due to Doppler shifting when changing from the lab to co-moving frame. The same algorithm is used for the inverse transformation.}
\label{fig:algorithm}
\end{figure} 

We detail our algorithm for transforming the Lorentz invariant intensity to account for Doppler shifting. When the Doppler shift is non-negligible, we must transform the argument of the Lorentz invariant intensity, as shown in (\ref{eq:intensity}). The intensity is stored at discrete frequencies $I(f_0), I(f_0 + \Delta f), ... , I(f_0 + (N_{\nu} - 1)\Delta f) $. In general, in transforming from the lab to co-moving frame via (\ref{eq:doppler}) the change in frequency $\Delta \nu = \nu_0 - \nu \neq k \Delta f$ for some integer $k$ i.e we must interpolate between intensity values in frequency space. We describe this algortihm in detail for the transformation from lab to co-moving frame. However, our method is identical for the inverse transformation.

For each discrete $f_{m}$ on the frequency grid in the co-moving frame we define the frequency in the lab frame f which satisfies the transformation
\begin{equation}
f_m = \Gamma \left( 1 - \frac{\mathbf{v} \cdot \mathbf{n}}{c}\right) f.
\end{equation}
Because $f$ is not in general on the frequency grid, we interpolate the intensity at this frequency from the nearest points on the frequency grid in the lab frame. Using cubic interpolation,
\begin{align}
I(f) &= -\frac{w (w-1) (w-2)}{6} I_{n-1} \nonumber \\
&+ \frac{(w+1) (w-1) (w-2)}{2} I_{n} - \frac{(w+1) w (w-2)}{2} I_{n+1} \\ 
&+ \frac{(w+1) w (w-2)}{6} I_{n+2}, \nonumber
\end{align} 
where $I_k = I(f_k)$ is the intensity at the appropriate frequency point, $f_n$ is the largest frequency on the grid that is less than $f$ and $w = (f - f_n)/\Delta f$. We illustrate this mapping graphically in Fig \ref{fig:algorithm}. Having interpolated on the frequency grid, we use the Lorentz invariance property of the intensity (\ref{eq:intensity}) to complete the transformation. We use an analogous method when performing the inverse transformation from the co-moving to the lab frame.   

\label{lastpage}

\end{document}